\documentclass[11pt,a4paper,oneside]{article}
\usepackage{color}
\usepackage{graphicx}
\usepackage[T1]{fontenc}

\usepackage{url}
\usepackage{textcomp}
\usepackage{amsmath}
\usepackage{amsfonts}
\usepackage{amsthm}
\usepackage{amssymb}
\usepackage[english]{babel}
\usepackage{xcolor}
\usepackage{appendix}

\usepackage{algorithm} 
\let\savealgorithm\algorithm
\let\saveendalgorithm\endalgorithm
\def\algorithm{\vskip-\lastskip\vskip-12pt\savealgorithm}
\def\endalgorithm{\saveendalgorithm\vskip-12pt}
\usepackage{algorithmicx}
\usepackage{algpseudocode}

\def\code{\bgroup\begin{itemize}\item[]\footnotesize}
\def\endcode{\end{itemize}\egroup}

\title{Manifold learning for brain connectivity}

\author{F\'elix Renard$^1$, Christian Heinrich$^2$, Marine Bouthillon$^2$, \\ Maleka Schenck$^{3,4}$, Francis Schneider$^{3,4,5}$, St\'ephane Kremer$^{2,6}$, Sophie Achard$^1$}

\date{\small $^1$ Univ. Grenoble Alpes, CNRS, Inria, Grenoble INP, LJK, 38000 Grenoble, France \\ $^2$ iCube, Universit\' e de Strasbourg, CNRS, 300 boulevard S. Brant, BP 10413, 67412 Illkirch Cedex, France \\ $^3$ Service de M\'edecine Intensive R\'eanimation, CHU de Strasbourg, France \\ $^4$ Facult\'e de M\'edecine FMTS, Strasbourg, France \\ $^5$ U1121, Universit\'e de Strasbourg, France \\ $^6$ Imagerie 2, CHU de Strasbourg, Universit\'e de Strasbourg, France}

\begin{document}

\maketitle

\textbf{keywords:} Graphs, machine learning, connectomes, hub disruption index

\begin{abstract}
Human brain connectome studies aim at extracting and analyzing relevant features associated to pathologies of interest. Usually this consists in modeling the brain connectome as a graph and in using graph metrics as features. A fine brain description requires graph metrics computation at the node level. Given the relatively reduced number of patients in standard cohorts, such data analysis problems fall in the high-dimension low sample size framework. In this context, our goal is to provide a machine learning technique that exhibits flexibility, gives the investigator grip on the features and covariates, allows visualization and exploration, and yields insight into the data and the biological phenomena at stake. The retained approach is dimension reduction in a manifold learning methodology, the originality lying in that one (or several) reduced variables be chosen by the investigator. The proposed method is illustrated on two studies, the first one addressing comatose patients, the second one addressing young versus elderly population comparison. The method sheds light on the graph metrics and underlying neurobiological phenomena.

\end{abstract}

\section{Introduction}

Brain modeling and understanding is a very active field of research involving different disciplines, such as neuroscience, image and signal processing, statistics, physics, and biology. These last years, neuroimaging modalities have been developed to explore the brain for both structural and functional features. It is now recognized that these images are providing very promising noninvasive observations of the brain \cite{mwangi2014review,bullmore.2009.1,richiardi2013}. One consequence of the availability of such massive datasets is the need to develop more and more sophisticated models to unravel the possible alteration of brains due to the impact of different pathologies. In this context, representing the brain as a global system is capital. This may be achieved using a \textit{network} \cite{bullmore2009}. A brain network is a graph where nodes correspond to specific regions and edges describe interactions and links between those regions. Different kinds of links and interactions may be of interest. Anatomical tracts are identified using diffusion imaging \cite{sporns2005} and used in anatomical connectivity studies, where the whole set of links is called an \textit{anatomical connectome}. Functional interactions are identified in functional imaging studies, whether in resting-state or in task-performing \cite{rosazza2011resting,Fallani2014}, and used in functional connectivity studies. The whole set of functional links is called a \textit{functional connectome}. In the functional case, brain networks are unique in encapsulating both spatial and temporal information in a single model. This model has attracted lots of attention these last twenty years by providing both very intuitive and spatial maps of brain networks. 

Brain networks can be quantified using graph metrics such as minimum path length, clustering \cite{watts1998}, global and local efficiency \cite{latora01}, modularity \cite{newman06}, and assortativity \cite{newman02}, among others. As these metrics are associated to specific network features, it is often possible to find the appropriate metrics to use given specific neuroscience hypotheses of the study. For the study of brain disorders, these metrics have been used in order to extract biomarkers for pathologies such as for example Alzheimer's disease \cite{supekar2008}, schizophrenia \cite{lynall2010}, and multiple sclerosis \cite{Filippi2014}. Extracting quantitative parameters of brain networks is compulsory to conduct any statistical analysis. In this framework, statistical and machine learning approaches on graph metrics on all nodes allow the quantification of differences between groups \cite{richiardi2013}.

 For any dataset, any graph metric can be computed either at the global level with one value for an entire network or at the nodal level with one value for each node and a vector of values for the entire network. It has already been shown that global values may not discriminate two groups of subjects \cite{achard2012}, which shows their limits as biomarkers. Few attempts have been made to use directly distances between networks such as the edit distance \cite{mokhtari2012}, or network similarities \cite{mheich2017siminet}. However, nodal level approaches are challenging since hundreds of brain areas can be extracted whereas  the number of subjects is generally small. This corresponds to the High Dimension Low Sample Size (HDLSS) configuration and falls under the curse of dimensionality \cite{bellman61}. In particular, standard classification and regression algorithms are not robust anymore in such a context (chapter 2 section 5 and chapter 18 of \cite{friedman2001}). 

Dimension reduction techniques tackle curse of dimensionality issues \cite{friedman2001}. In this framework, feature selection, where a subset of the original variables is considered, and feature extraction, where the original variables are transformed to a smaller set, may be envisaged \cite{webb02}. We resort here to the ISOMAP methodology, which is a well-known nonlinear feature extraction algorithm generalizing Principal Component Analysis dimension reduction \cite{tenenbaum2000,huo2007}. ISOMAP may be seen as a manifold learning approach, where the degrees of freedom of the data are captured by the latent variables, and where the structure of points in the latent space (the reduced space) mimics the structure of data in the original space. Nevertheless, ISOMAP raises two issues: interpreting the latent variables and determining the effect a change in the latent variables incurs in the data space, that is the corresponding changes in brain networks and the underlying neuroscience hypotheses at stake in the case of the present study. 

Dimension reduction is not new in the field of brain connectivity studies. Several methods have been proposed to extract nodal features at the level of brain regions. Using the Hub Disruption Index (the $\kappa$ index) to analyze a set of brain networks may be considered as a feature extraction approach: this is a user-defined transformation of the original space to a 1D latent space \cite{achard2012}. 
Principal Component Analysis (PCA) is applied in \cite{robinson2010}  on graph metrics vectors representing brains at nodal level. We proposed in \cite{renard2012} to use kernel PCA,  a nonlinear version of PCA. Besides, interpreting latent variables may be addressed by correlating the reduced space with clinical data \cite{gerber2010}. Covariates may also be mapped or regressed on the reduced space as proposed in \cite{aljabar2012}, thus shedding light on latent variables. 

The objective of this article is to integrate all features cited above in one method: working at the nodal level, applying dimension reduction techniques, and mapping covariates to ease interpretation. In addition, a new methodology is proposed to incorporate interesting networks features already identified in specific datasets directly in the manifold learning approach.   

This paper is focusing on two already published datasets. The first one consists in fMRI datasets on 20 healthy controls and 17 coma patients from Achard \textit{et al.} \cite{achard2012}. The second one is based on \cite{achard.2007.1} where 15 young healthy subjects and 11 elderly healthy subjects were scanned using resting state fMRI. Our first experiment compares  data driven approaches such as Linear Discriminant Analysis (LDA) and Random Forests (RF) to an ad hoc description such as the hub disruption index $\kappa$. This allows to compare classical machine learning approaches where the interpretability of the results is often difficult with approaches resorting to descriptors constructed using neuroscientific hypotheses. The second experiment consists in constructing a data-driven manifold, ISOMAP, using the graph metrics as features. ISOMAP is providing a compact representation of brain connectomes in a reduced space where it is straightforward to map the available covariates. In addition, we may interpret changes in connectomes by regressing  covariables like $\kappa$ on the reduced space using latent variables. 

This representation allows a visualization of each subject relatively to the whole population, which is crucial in clinical studies for example in order to better understand brain changes for each specific subject. Besides, $\kappa$ has been shown to be both a meaningful descriptor and a good classifying feature for brain connectomes of coma patients. Therefore, we propose a new method based on a covariate constrained manifold learning (CCML) using $\kappa$ as an input of ISOMAP. This allows us to propose a new generative model based on our new data representation, to better predict the evolution of each patient given the evolution of covariables.  

\section{Materials and methods}

\subsection{Resting state fMRI data}
\subsubsection{Comatose study}
The data were acquired in a previous study aimed at characterizing resting state connectivity brain networks for patients with consciousness disorders. The description of the data and results is reported in \cite{achard2012}. 
The patients were scanned a few days after major acute brain injury, when sedative drug withdrawal allowed for spontaneous ventilation. Therefore, all patients were spontaneously ventilating and could be safely scanned at the time of fMRI. The causes of coma are patient-dependent: 12 had cardiac and respiratory arrest due to various causes; 2 had a gaseous cerebrovascular embolism; 2 had hypoglycemia; and 1 had extracranial artery dissection. A total of twenty-five patients were scanned  (age range, 21-82 y; 9 men). Data on eight patients were subsequently excluded because of unacceptable degrees of head movement. The coma severity for each patient was clinically assessed using the 62 items of the WHIM scale: scores range from 0, meaning deep coma, to 62, meaning full recovery. Six months after the onset of coma, 3 patients had totally recovered, 9 patients had died, and 5 patients remained in a persistent vegetative state. The normal control group is composed of 20 healthy volunteers matched for sex (11 men) and approximately for age
(range, 25-51 y) to the group of patients. This study was approved by the
local Research Ethics Committee of the Faculty of Health Sciences of Strasbourg on October 24, 2008 (CPP 08/53) and by the relevant healthcare authorities. Written informed consent was obtained directly from the healthy
volunteers and from the next of kin for each of the patients. 
Resting-state data were acquired for each subject using gradient echo planar imaging technique with a 1.5-T MR scanner (Avanto; Siemens, Erlangen, Germany) with the following parameters: relaxation time = 3 s, echo time = 50 ms, isotropic voxel size = 4 x 4 x 4 $mm^3$, 405 images, and 32 axial slices covering the entire cortex. The preprocessing of the data is detailed in our previous study \cite{achard2012}. 

\subsubsection{Young and elderly study}

The data used in this study have already been analyzed
in two papers \cite{achard.2007.1} and \cite{meunier2009}. The goal of these papers was to identify the changes in brain connectomes for elderly subjects in terms of topological organization of brain graphs. The data consist of 15 young subjects aged 18-33 years, mean age=24 and 11 elderly subjects aged 62-76 years. Each subject was scanned using resting-state fMRI as described in \cite{achard.2007.1} (Wolfson
Brain Imaging Centre, Cambridge, UK). For each dataset, a total of 512 volumes was avalaible with number of slices, 21 (interleaved); slice thickness, 4 mm; interslice gap, 1 mm; matrix size, 64 3 64; flip angle,
908; repetition time (TR), 1100 ms; echo time, 27.5 ms; in-plane
resolution, 3.125 mm.

\subsection{Wavelet graph estimation}

Brain network graphs were determined following \cite{achard2012} for comatose study and \cite{achard.2007.1} for young and elderly study. For each subject, data were corrected for head motion and then coregistered with each subject's T1-weighted structural MRI. Each subject's structural MRI was non linearly registered with the Colin27 template image. The obtained deformation field image was used to map the fMRI datasets to the automated anatomical labeling (AAL) or to a customized parcellation image with 417 anatomically homogeneous size regions based on the AAL template image.
Regional mean time series were estimated by averaging the fMRI time series over all voxels in each parcel, weighted by the proportion of gray matter in each voxel of the segmented structural MRIs.
We estimated the correlations between wavelet coefficients of all possible pairs of the N = 90 or 417 cortical and subcortical fMRI time series extracted from each individual dataset.
For the coma, only scale 3 wavelet correlation matrices were considered. The corresponding frequency interval of the functional connectivity was 0.02-0.04 Hz. For the young and elderly, the wavelet scale considered corresponds to 0.06-0.11 Hz. 
To generate binary undirected graphs, a minimum spanning tree algorithm was applied to connect all parcels. The absolute wavelet correlation matrices were thresholded to retain 2.5 \% of all possible connections. Each subject was then represented by a graph with nodes corresponding to the same brain regions, and with the same number of edges.

\subsection{Graph metrics}

The objective is to extract differences between the two groups with respect to the topological organization of the graphs. Each graph is summarized by graph metrics computed at the nodal level. Three metrics are considered here: degree, global efficiency, and clustering \cite{bullmore.2009.1}. 

The degree is quantifying the number of edges belonging to one node. Let $G$ denote a graph with $G_{ij}=0$ when there is no edge between nodes $i$ and $j$, and  $G_{ij}=1$ when there is an edge between nodes $i$ and $j$.  The degree $D_i$ of node $i$ is computed as
\begin{equation}
D_{i} = \sum_{j \in G, j \neq i} G_{ij}.
\end{equation}

The global efficiency measures how the information is propagating in the whole network.
A random graph will have a global efficiency close to 1 for each node, and a regular graph will have a global efficiency close to 0 for each node.
The global efficiency $Eglob$ is defined as the inverse of the harmonic mean of the set of the minimum path lengths $L_{ij}$ between node $i$ and all other nodes $j$ in the graph:
\begin{equation}
Eglob_{i} = \frac{1}{N-1} \sum_{j \in G} \frac{1}{L_{ij} }
\end{equation}

Clustering is a local efficiency measure 
corresponding to information transfer in the immediate neighborhood of each node, defined as: \begin{equation}
Clust_i = \frac{1}{N_{G_i}(N_{G_i}-1)} \sum_{j,k \in G_i, \ j \neq k} \ \frac{1}{L_{jk}},
\end{equation} where $G_i$ is the subgraph of $G$ defined by the set of nodes that are the nearest neighbors of node $i$. A high value of clustering corresponds to highly connected neighbors of each node, whereas a low value means that the neighbors of each node are rather disconnected.

\begin{figure}[h]
    \begin{center}
    \includegraphics[scale=0.5]{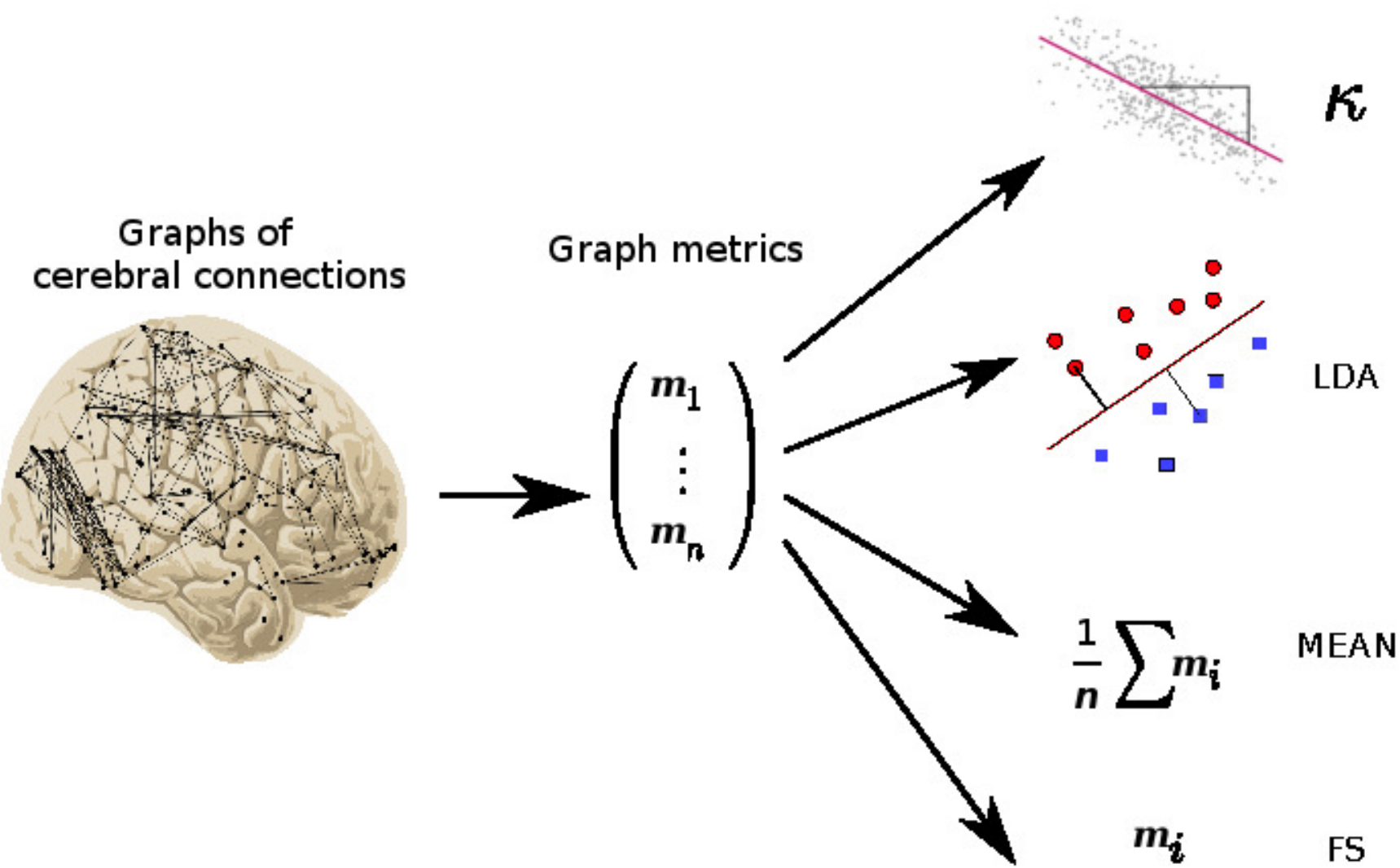}
    \end{center}
    \caption{General framework from graphs of cerebral connectomes to the different scalar features.}
    \label{schema}
\end{figure}

Each graph metric emphasizes a specific property at the nodal level. With a view to statistical comparison, several methods have already been developed, representing data in specific spaces. Each method aims at separating classes. Usually these methods are very general and can be applied without careful inspection of the data. We used here four different methods \cite{richiardi2011}: the $\kappa$ index resulting from a careful inspection of the data, mean over graph metrics (denoted here MEAN), LDA and Feature Selection (FS) by selecting the best feature based on a univariate statistical Student t-test. Like the $\kappa$ index, each of  these methods provides, for each patient, a scalar feature corresponding to particular property of the data. Figure \ref{schema} gives an illustration of the different methods.

\subsection{$\kappa$ index definition}
          
\begin{figure}[h]
    \begin{center}
    \includegraphics[scale=1.5]{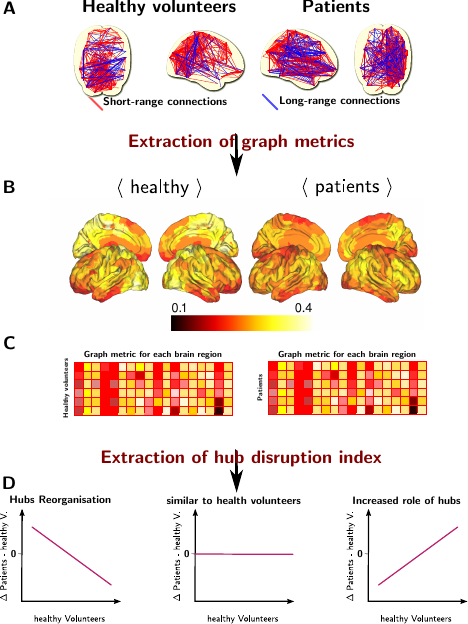}
    \end{center}
    \caption{Extraction of hub disruption index $\kappa$: A. brain connectomes inferred for each subjects; B. for each brain connectome, extraction of graph metrics for each region of the brain; C. matrix representation of the graph metrics where a row corresponds to a subject and a column corresponds to a brain region; D. computation of the hub disruption index by regressing the average of brain metrics of the difference of patients and average of healthy volunteers against the average of healthy volunteers. The hub disruption index corresponds to the slope coefficient. We give several illustrations following the sign of this coefficient.}
    \label{fig.kappa}
\end{figure}

In our previous study \cite{achard2012}, $\kappa$ was devised to compare graph metrics obtained on each node of a subject or of a group with reference group (see figure \ref{fig.kappa}). In  classical comparisons between a group of patients and a group of healthy volunteers, the reference is the group of healthy volunteers. In the present study, 
for a given graph metric and two groups, we first compute the average of this metric for each node over the group of healthy volunteers, denoted as the reference. Each subject is then summarized as a vector of values of dimension the number of nodes. Then, for each patient, $\kappa$ corresponds to the slope of the regression 
of a nodal graph metric between the given patient minus the reference and the reference. In order to give a simple interpretation of $\kappa$, we assume that the global graph metric computed as an average over the nodes is the same in both groups. A value of zero for $\kappa$ is showing that the graph metric obtained at the node level is the same for the patient and the reference. A positive value of $\kappa$ is indicating that the hubs and non-hubs of the patient in comparison to the reference are located on the same nodes. However, the values of the graph metrics are increased for the hubs and decreased for the non-hubs. Finally, when the value of $\kappa$ is negative, the hubs of the reference are no longer hubs of the patient, and the non-hubs of the reference are hubs for the patient.  In \cite{achard2012}, we showed that the $\kappa$ index is able to discriminate both groups (coma patients and healthy volunteers) while the global metric is unable to identify any significant difference. Instead of averaging the graph metrics, the $\kappa$ index is capturing a joint variation of the metrics computed for each node.   

\medskip

\subsection{Mean over the nodes (MEAN)}
For each graph metric, the mean over the nodes of the graph captures a global property of the network. These global metrics have been previously used to discriminate two populations of networks, for example for Alzheimer's disease \cite{supekar2008} and for schizophrenia \cite{lynall2010}. Such a coefficient can discriminate well two networks when their topologies are really different.
However, such  metrics do not take into account the specificity of the nodes. Indeed, when permuting the nodes of the graph, the global metric is not changed, but the hubs of the graph are not associated to the same nodes anymore. Therefore, a graph reorganization cannot be detected using such global metrics. 

\subsection{Linear discriminant analysis (LDA)}

LDA  \cite{fisher1936} is a classification method, aiming at identifying the linear projection optimally separating two groups. It can be considered as a gold standard for linear group discrimination. It is not specific to the analysis of networks.

LDA has been previously used for network discrimination in \cite{robinson2010}. This algorithm amounts to computing a scalar for each graph. However, there is no simple clinical interpretation of the discriminating parameter.

\subsection{Feature Selection (FS)}
As for LDA, FS determines the features yielding the best separation of the two groups. Several features may be used simultaneously. In order to establish a fair comparison with the other methods, we choose to extract the single feature yielding the best separation.
Several methods exist for FS. We choose univariate FS implemented in \cite{scikit-learn}.
An advantage of FS is that it is capturing discriminative features at the node level. As the selected features are extracted directly from the data, it is usally possible to derive a clinical interpretation. However, joint variations are not modeled and on the comatose study, FS is not able to yield results of the same quality as those obtained using $\kappa$. 

\subsection{Modeling populations of networks with manifold learning}

ISOMAP \cite{tenenbaum2000} is used as a manifold learning approach to describe population networks. We propose here an original approach based on ISOMAP, where we constrain one variable of the reduced space (the latent space) to correspond to a covariate.

\subsubsection{Manifold learning using ISOMAP}

ISOMAP devises a reduced dimension version of the original set of points. Interpoint distances in the reduced space reproduce as much as possible interpoint distances in the original space. Euclidean and geodesic distances are respectively used. Principal component analysis may be seen as a particular case of ISOMAP, where Euclidean distances are used in the original space, instead of geodesic distances. The reader is referred to \cite{tenenbaum2000} for details about the algorithm. 

In our case, the original data correspond to a vector of graph metrics for each subject, the dimension of the vector being the number of nodes times the number of metrics. For each analysis, only one metric is considered here. However, this method could be applied using jointly several metrics. Covariates may be regressed on the reduced space. In the present work, this was achieved using a classic radial basis function interpolation. 

\subsubsection{Covariate constrained manifold learning}
One drawback of manifold learning algorithms is the difficulty to interpret the reduced coordinates because they are usually meaningless. The original method proposed in this work consists in constraining one coordinate of the reduced space to correspond to a specific covariate. The other coordinates are left unconstrained, as in classical ISOMAP. Such a procedure requires special care regarding the optimization aspect. We apply a strategy proposed in \cite{brucher2008}, where points are introduced one by one.

Moreover, a scale factor $\alpha$ is considered for the axis corresponding to the covariate. This parameter, obtained by optimization, balances the scales of the different axes.

The reduced point $\tilde{\mathbf{x}}_i$ is defined by $ \tilde{\mathbf{x}}_i=[\alpha c_i ; \mathbf{x}_i ]^T $, where $c_i$ is  the chosen covariate and $\mathbf{x}_i$ are the other coordinates. The cost function $E$ is defined as:
\begin{equation}
    E = \sum_{i, i<j} \big( || \tilde{\mathbf{x}}_i - \tilde{\mathbf{x}}_j ||^2 - || \mathbf{y}_i - \mathbf{y}_j ||^2 \big) ^2
\end{equation}
For an incoming data point $i$, the cost function $E$ is optimized three times with regard to 1) $\mathbf{x}_i$  as $\displaystyle{\min_{\mathbf{x}_i} E }$ , 2) $\alpha$ as  $\displaystyle{\min_{\alpha} E }$ and 3) $\mathbf{x}_j$ for each point that has already been included as $\displaystyle{\min_{\{\mathbf{x}_j\}_{j=1...i-1}} E }$.

To facilitate optimization and to avoid possible local minima, instead of inserting the samples at random, we choose the sample to be incorporated next as the one with the largest geodesic distance to the samples already incorporated. Indeed, interpolation problems are always easier than extrapolation problems where greater uncertainty may occur.    
We initialize the procedure by taking the two samples with the largest geodesic distance.
The first two samples are used as landmarks of the border of the reduced space,
and the insertion of new samples will generate only small displacements of the already inserted samples.

The algorithm is described in Algorithm \ref{algo}.

\begin{algorithm}
\caption{-- covariate constrained manifold learning (CCML)}\label{algo}
\begin{algorithmic}
 \State \textit{Input -- dataset:} $N$ vectors (samples) $\{{\mathbf{y}_i}\}_{i=1..N}$ of a graph metric over $n$ graph nodes
 \State \textit{Result:} reduced space representation  $\{{\tilde{\mathbf{x}}_i}\}_{i=1..N}$  of the dataset, where the first coordinate of each $\tilde{\mathbf{x}_i}$ corresponds to the covariate.
 \State \textbf{Initialization:} select the two most distant samples
 \State \textbf{Determine their reduced coordinates} by minimizing $E$ with $\alpha = 1$
 \State \textbf{Update the scale $\alpha$} by minimizing $E$ wrt $\alpha$, $\mathbf{x}_i$ fixed, as $\displaystyle{\min_{\alpha} E}$.
 \While{All points are not included}
  \State \textbf{1) Select the most distant sample $\mathbf{y}_k$} to the already selected samples
  \State \textbf{2) Compute $\mathbf{x}_k$} by minimizing $E$ wrt $\mathbf{x}_k$ ($\alpha$ and other $\mathbf{x}_i$'s fixed) as $\displaystyle{\min_{\mathbf{x}_k} E}$.
  \State \textbf{3) Update the scale $\alpha$} by minimizing  $E$ wrt $\alpha$ ($\mathbf{x}_i$'s fixed) as $\displaystyle{\min_{\alpha} E }$.
  \State \textbf{4) Update $\mathbf{x}_j$'s of samples already included} by minimizing $E$ as $\displaystyle{\min_{\{\mathbf{x}_j\}_{j=1...k-1}} E }$.
 \EndWhile
 \end{algorithmic}
\end{algorithm}

\subsection{Application: a generative model for the prediction of the evolution of a subject with regard to the evolution of a covariate}

From the obtained embedding, a generative model 
\begin{equation}
\widehat{\mathbf{y}} = f(\tilde{\mathbf{x}})
\end{equation} can be devised, where $\widehat{\mathbf{y}}$ is a vector in the original space (the connectome space), $\tilde{\mathbf{x}}$ is a vector from the manifold embedding, and $f$ is a regression function.
MARS \cite{friedman91} is chosen for the regression function $f$ for its nice properties (one regression for each coordinate of $f$, i.e. $n$ regressions): locally linear and globally nonlinear. 
The parameters of $f$ can be determined using the dataset $\{{\mathbf{y}_i}\}_{i=1..N}$ and the corresponding reduced vectors $\{{\tilde{\mathbf{x}}_i}\}_{i=1..N}$ using equation:
\begin{equation}
\mathbf{y}_i = f(\tilde{\mathbf{x}}_i) + \boldsymbol{\epsilon}_i = \widehat{\mathbf{y}}_i + \boldsymbol{\epsilon}_i,
\end{equation} where $\boldsymbol{\epsilon_i}$ is the residual between a sample and its prediction $\widehat{\mathbf{y}}_i$. The residuals allow to evaluate the accuracy of the regression function.

This kind of model is not original, PCA being the most well known case where the model is defined as $\mathbf{y} = \mathbf{A \ \tilde{x}} + \boldsymbol{\epsilon}$, see e.g. \cite{lawrence2004,sfikas2016} for references. Such a generative model used in the CCML framework allows to determine changes in the original space (the connectome space) generated by a displacement in the reduced space, for example along the covariate axis.

\section{Results}

The different algorithms have been implemented in the Python language using the scikit learn toolbox \cite{scikit-learn}. When left unspecified, coma data are used. The use of the young and elderly data is explicitely stated.

\subsection{Local analysis using dimension reduction} \label{res_local_analysis}
Permutation tests are performed on the $\kappa$ index and on the three other measures (LDA, FS, MEAN) to assess the ability of those four metrics to discriminate two populations. 
More precisely, for each coefficient separately, the difference of the means of the two populations is determined for the observed populations.
The labels of the samples are then shuffled and the difference of the means of the shuffled two populations is determined.
This latter step is performed $10^4$ times. It avoids to make any assumption on the distribution of the statistic.
Simultaneously the correlations between the observed $\kappa$ index and the other coefficients are estimated.

\begin{table}[h]
    \caption{\label{scores} P-value of permutation tests comparing the mean of the two groups ($10^4$ permutations, which bounds the p-values).
    The correlation scores are estimated between the $\kappa$ index and the three other measures ($LDA$, $FS$, and $MEAN$).}
    \begin{center}
    \begin{tabular}{||c||*{6}{c|}|}
        \hline
        \hline
        Coefficients & $Eglob$ & corr.  &  $Clust$ & corr.  & $D$    & corr.  \\ 
                     & (p-value)         &              & (p-value)   &              & (p-value) &            \\ \hline \hline
        $\kappa$   & $<10^{-4}$  &      &  $<10^{-4}$   &        &   $<10^{-4}$     &       \\ \hline
        $LDA$   & $<10^{-4}$ &  $0.88$ & $<10^{-4}$ & $0.87$   &   $<10^{-4}$ & $0.88$  \\ \hline
        $FS$    & $<10^{-4}$  &  $0.6$  &  $<10^{-4}$ & $-0.66$ &   $<10^{-4}$ & $0.6$   \\ \hline
        $MEAN$  & $ 0.58 $             &  $0.25$ &  $0.43$              & $-0.19$  &                      &         \\ \hline \hline
    \end{tabular}
    \end{center}
\end{table}

The results corresponding to the different methods aiming at discriminating the two groups (control and coma) are given in Table \ref{scores}. As expected, the machine learning algorithms (ie, $LDA$, $FS$, and $MEAN$) show good performances in separating the two groups for different graph metrics. This is consistent with the fact that these methods have been tailored to classify the two groups. The results with the $\kappa$ index show similar performances in separating the two groups. The large correlations between machine learning algorithms on the one hand and the graph metric $\kappa$ on the other hand show retrospectively that similar performances were to be expected.

Besides, a strong relationship can be observed between $\kappa$ and LDA (correlation scores greater than $0.87$ for each metric). The FS correlation scores are lower than the LDA correlation ones. The difference between the two methods is that LDA considers a linear combination of features, with a global perspective, whereas FS selects one feature and acts locally. Since $\kappa$ reflects a global reorganization of the brain, it is expected that the correlation score of LDA be greater than the FS one. Finally, MEAN scores reveal that this measure is not appropriate in this study.

\subsection{Standard ISOMAP manifold learning and the $\kappa$ index}\label{manifold_kappa}

\begin{figure}[h]
    \begin{center}
    \includegraphics[scale=0.15]{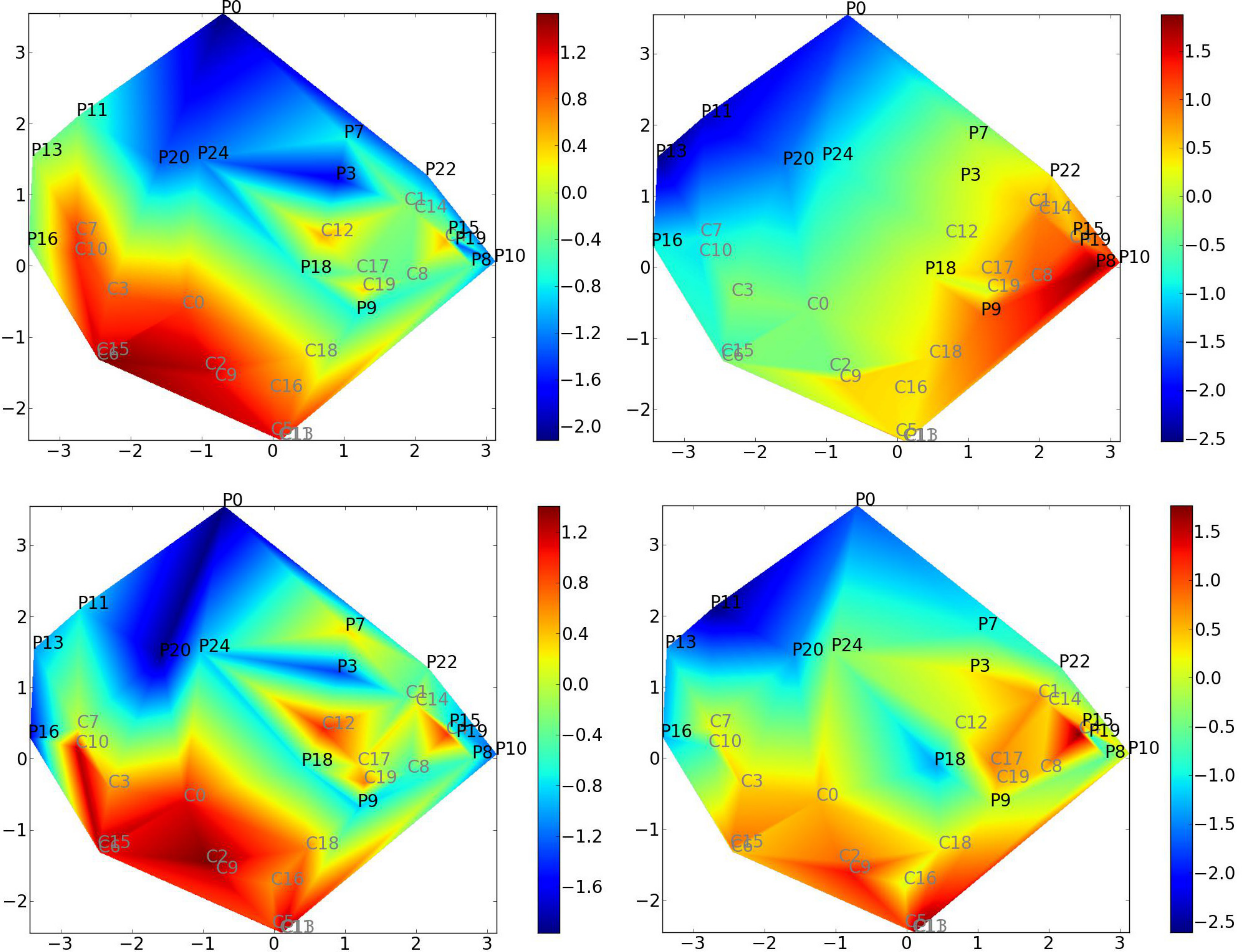}
    \end{center}
    \caption{Standard ISOMAP reduced space representation of the original dataset. $Pi$: (comatose) patient $\# i$ ; $Cj$: control $\#j$. Covariates are mapped onto the reduced space (covariate value is color-coded).  Top left: $\kappa$ index mapping; top right: MEAN mapping; bottom left: LDA mapping; bottom right: FS mapping.}
    \label{GE}
\end{figure}

In this section, the goal is to link the reduced space obtained by manifold learning to different covariates such as the $\kappa$ index. We want to assess whether a given covariate varies smoothly across the reduced space, and is therefore predictible using this space. 

Figure \ref{GE} represents the reduced space obtained using standard ISOMAP, as opposed to using CCML. The values of the different covariates are color-coded. The reduced space representation allows to separate both populations. Besides, by visual inspection of the color-coded maps, it appears that those regression maps are capturing features corresponding to $\kappa$ and MEAN. 
In order to quantify these visual observations, covariates are regressed on the reduced space. The root mean square error (RMSE) and the maximum error $M$ are determined in a leave-one-out procedure. The results are given in Table~\ref{reg}. It can be noted that the MEAN strategy gives the same values for the graph metric degree D for all graphs since the number of edges is set to be the same for all graphs.

\begin{table}[h]
    
    \begin{center}
    \begin{tabular}{||c||*{6}{c|}|}
        \hline \hline
        Covariate & $Eglob$ & $Eglob$ & $Clust$ &  $Clust$ & $D$ & $D$ \\  & RMSE & M & RMSE & M & RMSE & M  \\ \hline \hline
        $\kappa$        & 0.51          & 2.44         &  0.68           &  2.59           &  0.26          &   2.14        \\ \hline
        $LDA$        & 0.97          & 7.97         &  0.9            &  5.44           &  0.66          &   4.13        \\ \hline
        $FS$         & 0.64          & 2.2          &  1.1            &  9.12           &  0.83          &   7.49        \\ \hline
        $MEAN$       & 0.15          & 0.73         &  1.02           &  5.44           &                &               \\ \hline\hline
    \end{tabular}
    \caption{\label{reg} Assessment of the regression of covariates on the reduced space. Three different reduced spaces are at stake, one for each graph metric. Root mean square error (RMSE) and maximal error $M$ are displayed. The MEAN strategy  is not relevant for the degree D since the degrees D for all graphs are equal (the number of edges is set to be the same for all graphs). }
    \end{center}
\end{table}

In this table, the lower the values, the better the adequacy with the reduced space. It appears that $\kappa$ is the best choice across all metrics, except for $Eglob$ where it is outperformed by MEAN. In the case of $Eglob$, this suggests that both $\kappa$ and MEAN scores correspond to degrees of freedom of the intrinsic manifold of the functional connectivity graphs.

\begin{figure}[h]
    \begin{center}
    \includegraphics[scale=0.2]{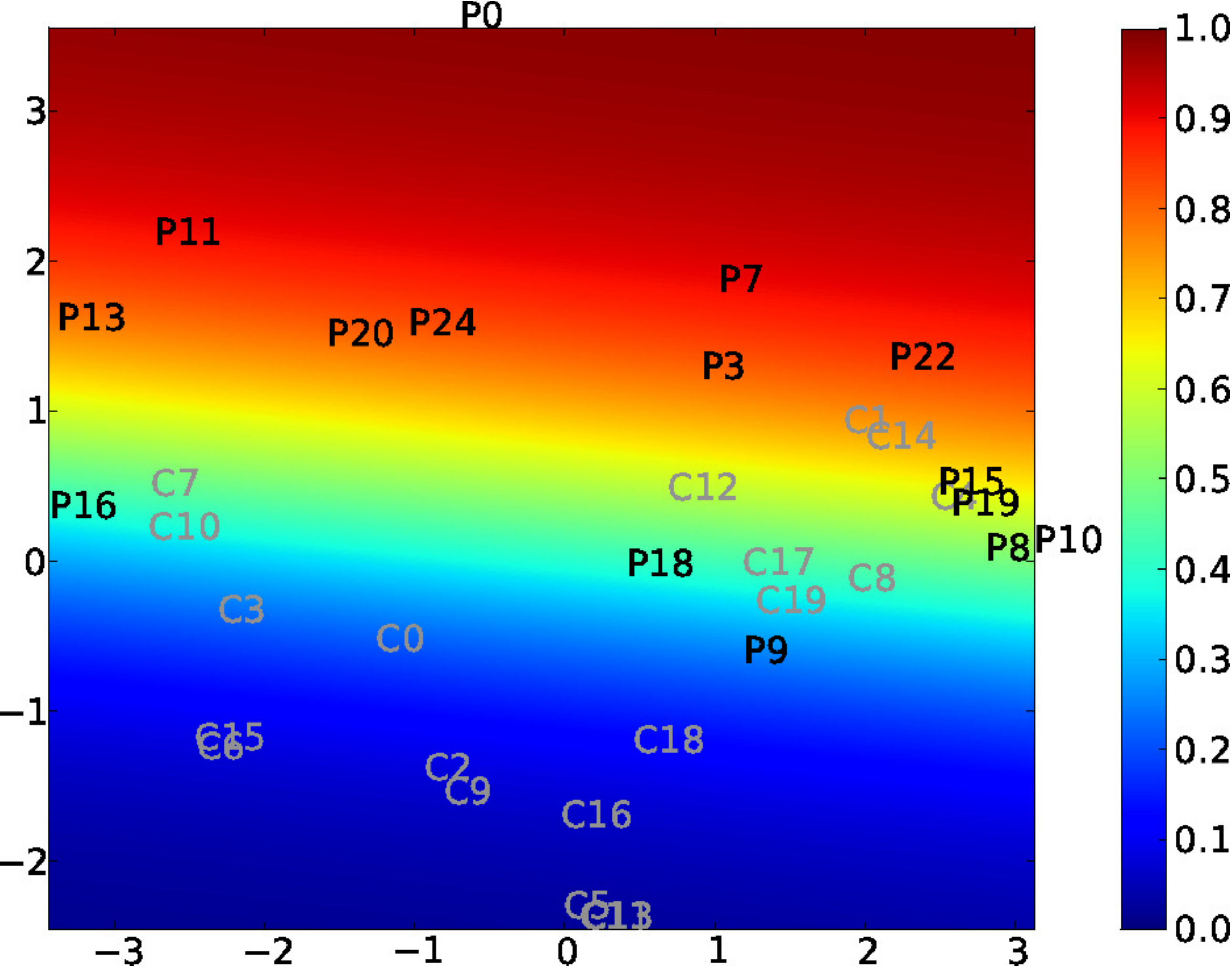}
    \end{center}
    \caption{Logistic regression using reduced space stemming from standard ISOMAP. The color codes the probability of belonging to the comatose class.} 
    \label{logistique}
\end{figure}

Figure \ref{logistique} displays the probability of belonging to a specific class computed using logistic regression on the reduced space stemming from standard ISOMAP. The probability of belonging to the comatose class is color-coded in Figure \ref{logistique}. 

This probability estimation using logistic regression is compared 
with covariates such as $\kappa$ or MEAN, in Table~\ref{score_log}. A high correlation score is observed between  $\kappa$ and logistic regression probablity. 
The correlation score between the MEAN coefficient and the probabilistic mapping is lower than the one with $\kappa$ as expected.

\begin{table}[h]
    \begin{center}
    \begin{tabular}{||c||*{3}{c|}|}
        \hline \hline
        Coefficients & $Eglob$         & $Clus$        & $D$          \\ \hline \hline
        $\kappa$        & 0.87            &  0.87         &  0.81         \\ \hline
        $MEAN$       & 0.55            &  0.42         &               \\ \hline\hline
    \end{tabular}
    \caption{\label{score_log} Correlation scores  between the probabilistic mapping and the different coefficients ($\kappa$ index and MEAN measure). A high correlation score of the $\kappa$ index indicates a good fitting between the reduced space representation and the classification of the two groups. }
    \end{center}
\end{table}

Taken together, these observations demonstrate the importance of $\kappa$ in the classification of these populations.
Obviously, for the special case of global efficiency metric, the MEAN score describes correctly the reduced space, but does not correspond to the classification pattern.

\subsection{Covariate constrained manifold learning}\label{new_iso}

\subsubsection{Comatose population}

 First we evaluate the convergence of the optimization problem (Algorithm \ref{algo}). To assess the difficulty of the optimization problem, we ran it with $500$ random initializations. Only $63 \%$ of the runs converged to the same solution, whereas $37 \%$ of the runs converged to a local (worse) optimum. It thus appears that our initialization procedure addresses the local optimum issue. Nevertheless, this optimization problem would probably deserve further investigations which are out of the scope of this paper. 

\begin{figure}[h]
    \begin{center}
    \includegraphics[scale=0.15]{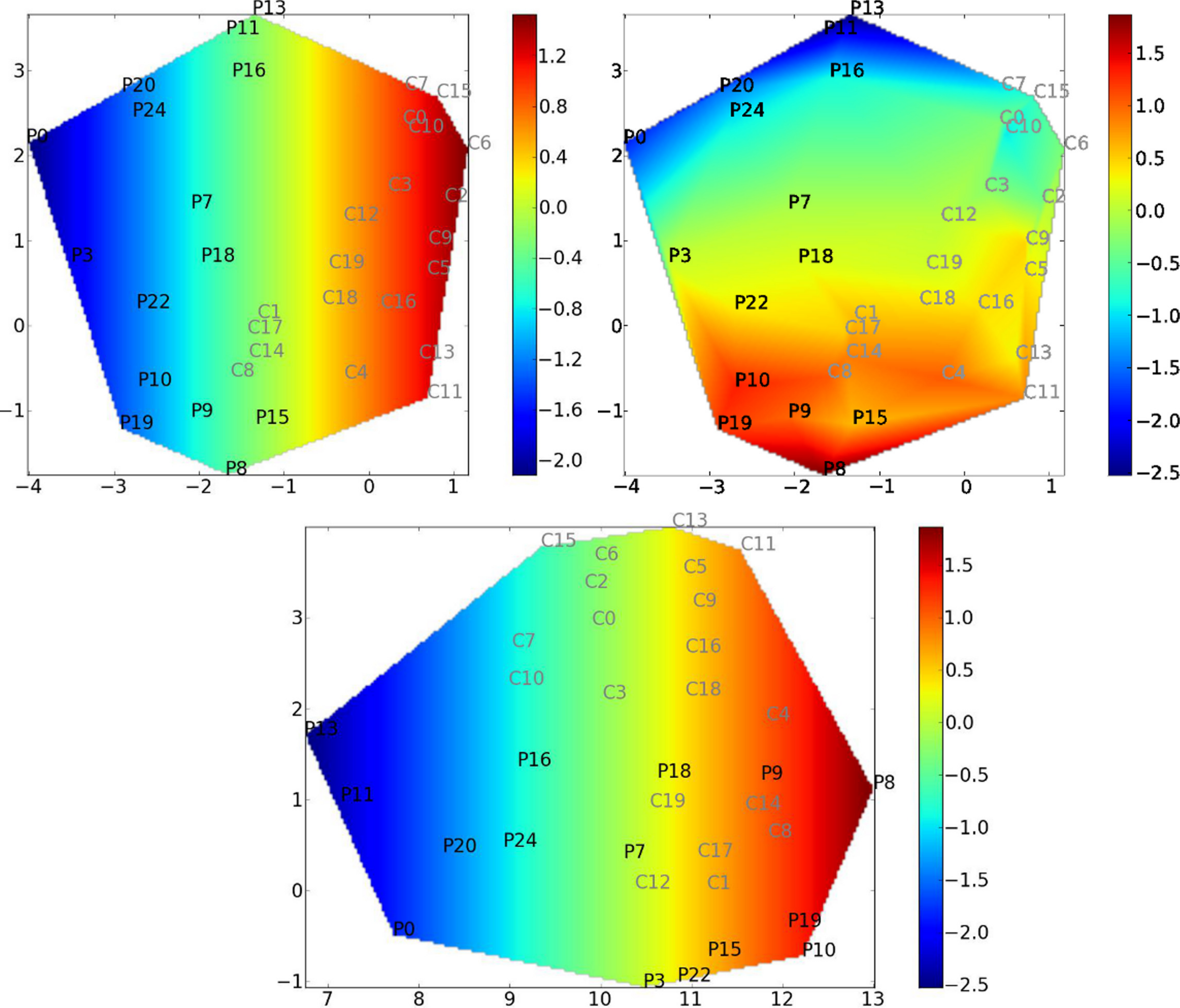}
    \end{center}
    \caption{Covariate mapping onto the reduced space given by our method CCML. The reduced space is computed using $E_{glob}$ as a graph metric (the $y_i$'s). Covariate value is color-coded. For each subfigure, the coordinates correspond to $[\alpha c_i ; x_i]^T$, where $c_i$ is the constrained variable and $x_i$ the free parameter. Top left: $\kappa$ mapping with a $\kappa$-constrained reduced space, Top right: MEAN mapping with a $\kappa$-constrained reduced space; Bottom : MEAN mapping with a MEAN-constrained reduced space. As expected, we can observe that the mappings correlate well with the first coordinate by construction (top left and bottom). It is also clear that using a $\kappa$-constrained reduced space is facilitating the discrimination between the two populations. Indeed, the controls and patients are not covering the same part of the reduced space. On the contrary, as expected using the MEAN-constrained reduced space, the method is not providing a very clear discrimination between patients and controls. Especially, patients 9 and 18 are very close to controls. Finally, the top right figure is showing a correlation between the second coordinate of CCML and the MEAN.}
    \label{new_iso_fig}
\end{figure}

In Figure \ref{new_iso_fig}, we display the reduced space corresponding to our new manifold learning algorithm. We can observe that the two populations are well discriminated in the case of $\kappa$, but not for the MEAN coefficient.

We observe the strong interaction between $\kappa$ and the MEAN in the top right of Figure \ref{new_iso_fig}, where the reduced space on one axis is  $\kappa$ and the mapping corresponds to the MEAN. To quantify this, in the case of  $\kappa$, we estimate the correlation between the second coordinate and the MEAN score. The obtained score equals to $0.92$, which confirms the intrinsic relationship between  $\kappa$ and the MEAN coefficient.

\subsubsection{Elderly and young population}

The elderly and young groups are investigated in this section.

\begin{figure}[h]
    \begin{center}
    \includegraphics[scale=0.15]{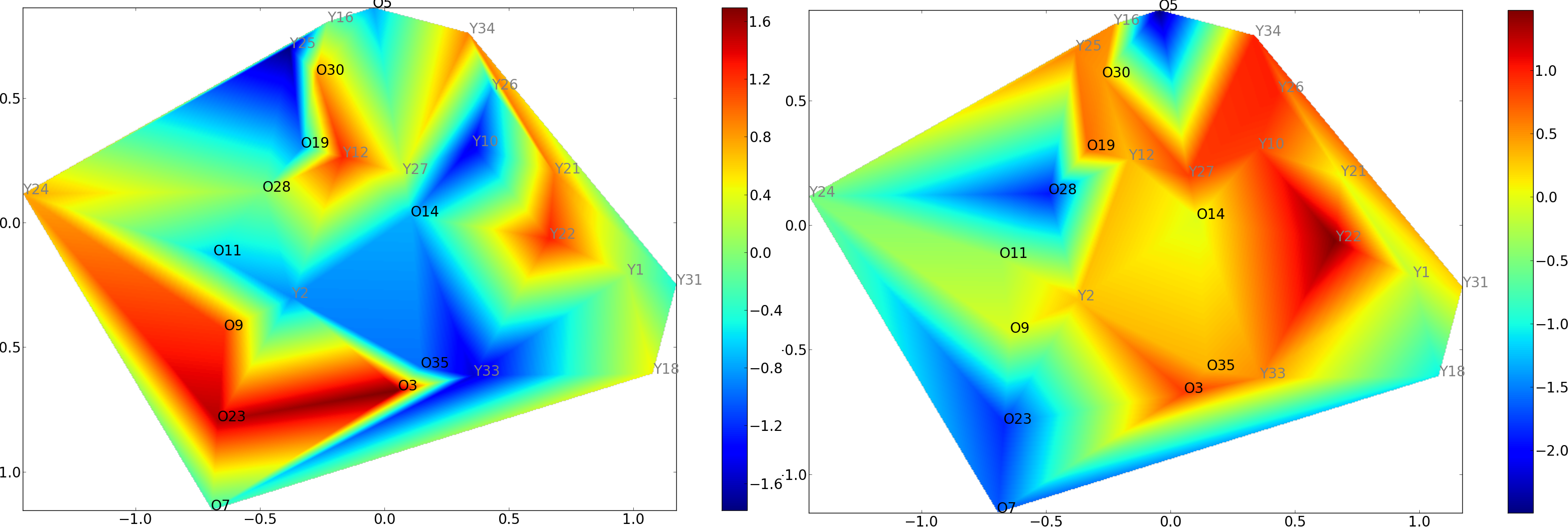}
    \end{center}
    \caption{Left: $\kappa$ mapping with the standard ISOMAP reduced space, Right: MEAN mapping with the same reduced space. The old controls (resp. young controls) are labeled O (resp. Y). For these groups, the $\kappa$ index cannot discriminate the two groups. However the MEAN index behaves better for the discrimination between the two groups. In each case, the interpretation is  complex since the mapping of the covariate is not linear.}
    \label{oy_mani_natural_fig}
\end{figure}

\begin{figure}[h]
    \begin{center}
    \includegraphics[scale=0.15]{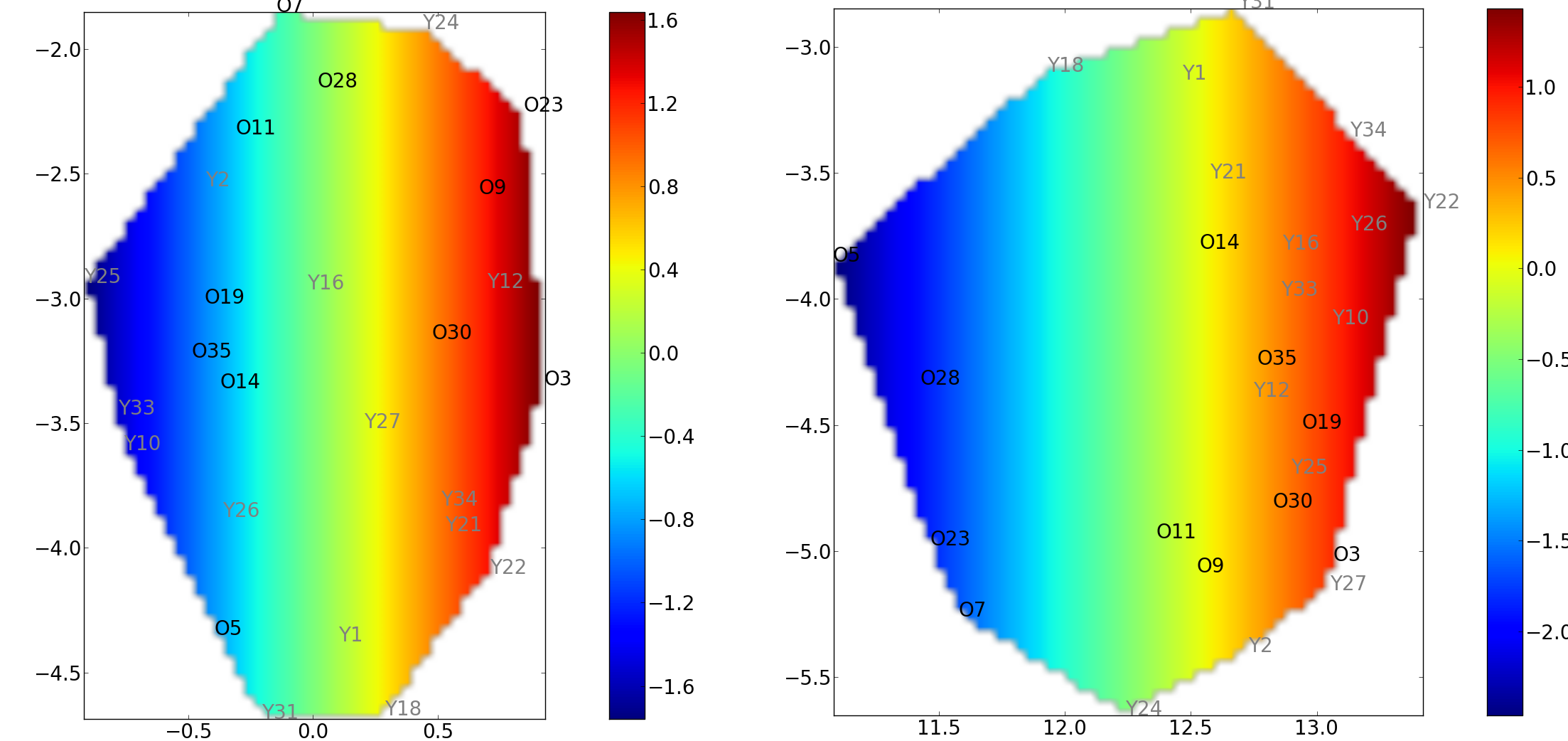}
    \end{center}
    \caption{CCML approach. Left: $\kappa$ mapping with the $\kappa$-constrained reduced space, Right: MEAN mapping with the MEAN-constrained reduced space. The old controls (resp. young controls) are labeled O (resp. Y). Two manifolds (one for each constraint) have been determined. However only the MEAN-constrained manifold  discriminates both groups.}
    \label{oy_mani_fig_1}
\end{figure}

In Figure \ref{oy_mani_natural_fig}, the manifold obtained by standard ISOMAP is displayed. It is interesting to highlight that the $\kappa$ index is not a pertinent feature to discriminate the old from the young, whereas the MEAN is a better discriminating feature. In both cases, the interpretation of the mapping is complex since it is not smooth.

In Figure \ref{oy_mani_fig_1}, results from CCML are displayed for the MEAN coefficient. We can observe that the MEAN mapping discriminates the two groups.

\subsection{Application: a generative model for the prediction of the evolution of a subject with regard to the evolution of a covariate}

Using the algorithm detailed in the last section, a map of the population is determined, with one of the reduced coordinates corresponding to a chosen covariate. To highlight the potential of the proposed method, we compute the evolution of a patient with regard to the evolution of a covariate. This allows gives insight into the  effect of the covariate on the patient. To perform this analysis, a regression is used to map the reduced space to the initial space (we used MARS regression \cite{friedman91}, coordinate-wise).

\begin{figure}[h]
    \begin{center}
    \includegraphics[scale=0.7]{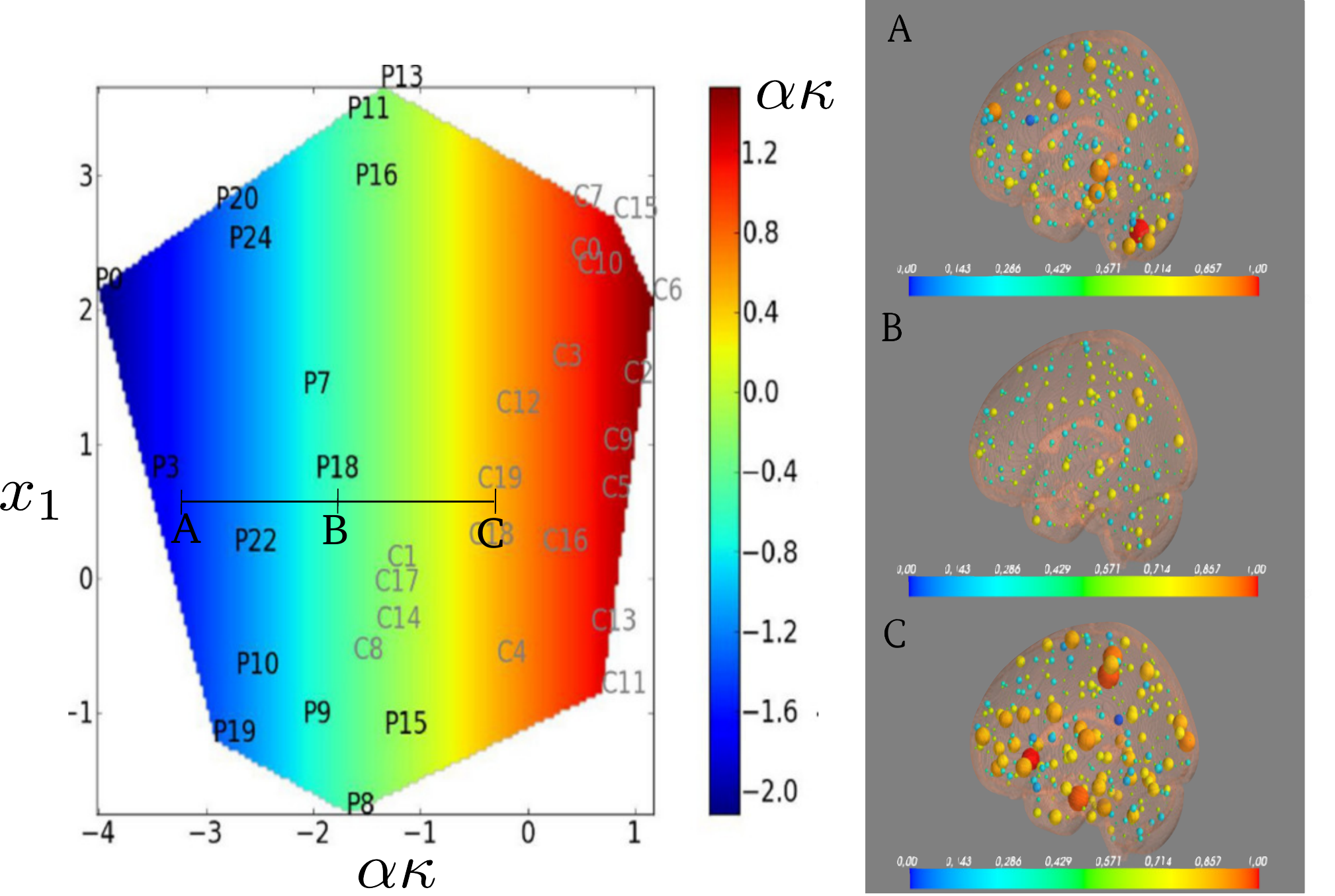}
    \end{center}
    \caption{Evolution of one patient along the covariate axis. We consider the patient P18 in state B in the reduced space (left part of the figure). We predict its evolution when the $\kappa$ index is decreased (point A) or increased (point C). Intuitively, point A and point C correspond respectively to a degradation and to an improvement of the health of the patient. Interestingly, these trends can be directly observed in the graph space for clinical insights (right part of the figure).  It can be noticed that the variations are not linear. }
    \label{patient_specific}
\end{figure}

This application is illustrated in Figure \ref{patient_specific}.

\section{Discussion}

\subsection{Assessment of graph metric descriptors}

The handcrafted design of graph metric descriptors is interesting since such descriptors carry straightforward physical meaning, like the $\kappa$ index for hub reorganization. However, in a classification framework, such new scalar descriptors may not be optimal. To assess the pertinence of a new \textit{ad hoc} graph metric descriptor for a classification task, it will be interesting to confront it to specific scalar coefficients used in standard classification algorithms (like LDA or FS), and examine if there is some correlation between the scalars at stake. 

\subsection{No free lunch for graph metric descriptors}

We hypothesize that there is no  best descriptor adapted to all datasets. The optimality depends on 1) the kind of the data and 2) the kind of question/task addressed. This idea is known as the "no free lunch theorem"  \cite{wolpert1996}: if an algorithm performs well on a certain class of problems then it has necessarily poorer performances on the set of all remaining problems. 

In the present study, we showed that the $\kappa$ index yields good classification performances in separating a comatose population from a healthy population. However the MEAN index better describes the groups of elderly and young people (see Fig.~\ref{GE}): for this dataset, the $\kappa$ index cannot separate the two groups, but the MEAN score can. It is interesting to notice that several descriptors can map correctly a population, while providing different information.

The "no free lunch theorem" also applies to manifold learning algorithms. The underlying question is the one of choosing an interpoint distance in the data space. A given interpoint distance will yield a specific structure of samples in the reduced space. Therefore, the retained interpoint distance chosen will depend on the final goal: mimicking the structure of the initial data points, enhancing class separation with a view to achieving better classification performances,  focusing on a specific property of the data, etc\dots The proposed algorithm CCML aims at mimicking the structure of the initial data points, and this to be done  using  explicitly a particular characteristic, chosen and imposed by the investigator.

\subsection{Manifold learning for brain graph analysis}

Manifold learning is well suited for brain graph analysis for several reasons.
Firstly, global descriptors of graph metrics represent an entire graph by a scalar value, which is generally ultimately insufficient to model correctly the complexity of a graph population. Manifold learning is better suited to capturing the complexity and variability of a given graph population, since more degrees of freedom are structuring the reduced space.

Secondly, connectomes have been studied for their capacity to represent the brain as a system and not merely as a juxtaposition of independent anatomical/functional regions. Classical statistical tests are not adapted to analyze joint variations between local descriptors of graph metrics since those tests assume independence between features. Brain graph manifold learning for comparing  groups of graphs  is promising because joint variations are accounted for. 

Thirdly, manifold learning may be turned to a generative model, when resorting to a mapping from the reduced space to the data space.

Brain graph manifold learning can be seen as a trade-off between global and local brain graph metrics analysis. In other words, manifold learning is considered as a model at the level of the group while preserving the information of the individuals. However this technique is hard to interpret by its own.
The addition of explicative covariables as proposed with the CCML method can provide an understandable and generative model of population with the possibility of focusing at the individual level \cite{costa2015searching}.

More generally, manifold learning can be an interesting solution for personalized and predictive medicine purposes.

\section{Conclusion}

The originality and contribution of this paper is the devising of a nonlinear manifold model of brain graph metrics. The essence of the approach is the capture of a metric through all nodes of a graph, across all graphs of an entire population: a population of graphs is represented by a population of vectors, each vector holding the variation of a metric through the nodes at stake. The population is then represented in a reduced space. This is to be opposed to the standard representation of a given brain graph by a mere scalar.

The proposed approach has several advantages. First and foremost, the data are represented with several degrees of freedom, corresponding to the dimensions of the reduced space. The structure of the original data set is captured by a compact representation.  This allows to account for the complex variability of populations of brain graphs. Secondly, such an approach naturally offers analysis of joint variations of those brain graph metrics. Besides, the investigator has the possibility to analyze the data at the population scale and simultaneously at the individual scale.

The investigation tool corresponding to the proposed approach allowed us to retrospectively assess the hub disruption index (HDI), denoted $\kappa$, and proposed in one of our former works. Earlier work showed that the HDI is a very good candidate for discriminating patients and controls in the case of coma. Retrospectively, its performances are here assessed in comparison with machine learning methods dedicated to linear group classification such as LDA. Besides yielding nice classification performances, the present study showed that an advantage of HDI, put in the perspective of a manifold model, is to give clinical clues related to the pathology mechanism.

We  observed strong relationships between scalar coefficients such as HDI and MEAN, and the coordinates of the manifold. It is important to notice that MEAN, which can separate groups in several  pathologies \cite{supekar2008,lynall2010}, is not able to discriminate the comatose patients from the normal population. However it brings additional information in terms of description of the population. The  manifold at stake shows that a scalar coefficient cannot capture the whole information encapsulated in the graphs. One interest of manifold learning, and more specifically our new proposed method, is its ability to reach a new level of interpretation of the brain graph metrics and the interaction between them.\\

\textbf{Author contributions}\\
This project was formulated by FR, CH and SA based on substantial data, analyses and experiments of FR, CH, MB, MS, FS, SK and SA. 
FR and SA formalised the model, implemented and ran the model;
FR, CH and SA wrote the manuscript.


\begin{thebibliography}{10}

\bibitem{achard.2007.1}
S.~Achard and E.~Bullmore.
\newblock Efficiency and cost of economical human brain functional networks.
\newblock {\em {PLoS} Computational Biology}, 3:{e}17, 2007.

\bibitem{achard2012}
Sophie Achard, Chantal Delon-Martin, Petra~E. V\'{e}rtes, F\'{e}lix Renard,
  Maleka Schenck, Francis Schneider, Christian Heinrich, St\'{e}phane Kremer,
  and Edward~T. Bullmore.
\newblock Hubs of brain functional networks are radically reorganized in
  comatose patients.
\newblock {\em Proceedings of the National Academy of Sciences},
  109(50):20608--20613, Nov. 2012.

\bibitem{aljabar2012}
P.~Aljabar, R.~Wolz, and D.~Rueckert.
\newblock Manifold learning for medical image registration, segmentation and
  classification.
\newblock In K.~Suzuki, editor, {\em Machine learning in computer-aided
  diagnosis: medical imaging intelligence and analysis}. IGI Global, 2012.

\bibitem{bellman61}
Richard~E. Bellman.
\newblock {\em {Adaptive control processes - A guided tour}}.
\newblock Princeton University Press, 1961.

\bibitem{brucher2008}
Matthieu Brucher, Christian Heinrich, Fabrice Heitz, and Jean{-}Paul Armspach.
\newblock A metric multidimensional scaling-based nonlinear manifold learning
  approach for unsupervised data reduction.
\newblock {\em {EURASIP} J. Adv. Sig. Proc.}, 2008, 2008.

\bibitem{bullmore.2009.1}
E.~Bullmore and O.~Sporns.
\newblock Complex brain networks: graph theoretical analysis of structural and
  functional systems.
\newblock {\em Nat Rev Neurosci}, 10(3):186--198, Mar 2009.

\bibitem{bullmore2009}
Ed~Bullmore and Olaf Sporns.
\newblock Complex brain networks: graph theoretical analysis of structural and
  functional systems.
\newblock {\em Nature Reviews Neuroscience}, 10(3):186--198, feb 2009.

\bibitem{costa2015searching}
Lilia Costa, Jim Smith, Thomas Nichols, James Cussens, Eugene~P Duff, Tamar~R
  Makin, et~al.
\newblock Searching multiregression dynamic models of resting-state fmri
  networks using integer programming.
\newblock {\em Bayesian Analysis}, 10(2):441--478, 2015.

\bibitem{Fallani2014}
Fabrizio De~Vico Fallani, Jonas Richiardi, Mario Chavez, and Sophie Achard.
\newblock Graph analysis of functional brain networks: practical issues in
  translational neuroscience.
\newblock {\em Philosophical Transactions of the Royal Society B: Biological
  Sciences}, 369(1653):20130521, 2014.

\bibitem{Filippi2014}
Massimo Filippi, Paola Valsasina, Sara Sala, Vittorio Martinelli, Angelo
  Ghezzi, Pierangelo Veggiotti, Andrea Falini, Giancarlo Comi, and Maria Rocca.
\newblock Abnormalities of the brain functional connectome in pediatric
  patients with multiple sclerosis.
\newblock {\em Neurology}, 82(10 Supplement), 2014.

\bibitem{fisher1936}
R.~A. Fisher.
\newblock The use of multiple measurements in taxonomic problems.
\newblock {\em Annals of Eugenics}, 7(2):179--188, 1936.

\bibitem{friedman91}
J.~Friedman.
\newblock Multivariate adaptive regression splines.
\newblock {\em Annals of Statistics}, 19(1):1--67, March 1991.

\bibitem{gerber2010}
Samuel Gerber, Tolga Tasdizen, P.~Thomas~Fletcher, Sarang Joshi, and Ross
  Whitaker.
\newblock Manifold modeling for brain population analysis.
\newblock {\em Medical Image Analysis}, 14(5):643--653, Oct. 2010.

\bibitem{friedman2001}
Trevor Hastie, Robert Tibshirani, and Jerome Friedman.
\newblock {\em The elements of statistical learning}.
\newblock Springer, 2001.

\bibitem{huo2007}
Xiaoming Huo, Xuelei~Sherry Ni, and Andrew~K Smith.
\newblock A survey of manifold-based learning methods.
\newblock {\em Recent advances in data mining of enterprise data}, pages
  691--745, 2007.

\bibitem{latora01}
V.~Latora and M.~Marchiori.
\newblock Efficient behavior of small-world networks.
\newblock {\em Physical Review Letters}, 87:198701, Oct 2001.

\bibitem{lawrence2004}
Neil~D Lawrence.
\newblock Gaussian process latent variable models for visualisation of high
  dimensional data.
\newblock In {\em Advances in neural information processing systems},
  volume~16, pages 329--336, 2004.

\bibitem{lynall2010}
Mary-Ellen Lynall, Danielle~S. Bassett, Robert Kerwin, Peter~J. McKenna,
  Manfred Kitzbichler, Ulrich Muller, and Ed~Bullmore.
\newblock Functional connectivity and brain networks in schizophrenia.
\newblock {\em The Journal of Neuroscience}, 30(28):9477--9487, 2010.

\bibitem{meunier2009}
David Meunier, Sophie Achard, Alexa Morcom, and Ed~Bullmore.
\newblock Age-related changes in modular organization of human brain functional
  networks.
\newblock {\em Neuroimage}, 44(3):715--723, 2009.

\bibitem{mheich2017siminet}
Ahmad Mheich, Mahmoud Hassan, Mohamad Khalil, Vincent Gripon, Olivier Dufor,
  and Fabrice Wendling.
\newblock Siminet: a novel method for quantifying brain network similarity.
\newblock {\em IEEE transactions on pattern analysis and machine intelligence},
  40(9):2238--2249, 2017.

\bibitem{mokhtari2012}
Fatemeh Mokhtari and Gholam-Ali Hossein-Zadeh.
\newblock Decoding brain states using backward edge elimination and graph
  kernels in f{\sc mri} connectivity networks.
\newblock {\em Journal of neuroscience methods}, 212(2):259--268, 1 2013.

\bibitem{mwangi2014review}
Benson Mwangi, Tian~Siva Tian, and Jair~C Soares.
\newblock A review of feature reduction techniques in neuroimaging.
\newblock {\em Neuroinformatics}, 12(2):229--244, 2014.

\bibitem{newman02}
M.~E.~J. Newman.
\newblock Assortative mixing in networks.
\newblock {\em Physical Review Letters}, 89:208701, Oct 2002.

\bibitem{newman06}
M.~E.~J. Newman.
\newblock Modularity and community structure in networks.
\newblock {\em Proceedings of the National Academy of Sciences},
  103(23):8577--8582, 2006.

\bibitem{scikit-learn}
F.~Pedregosa, G.~Varoquaux, A.~Gramfort, V.~Michel, B.~Thirion, O.~Grisel,
  M.~Blondel, P.~Prettenhofer, R.~Weiss, V.~Dubourg, J.~Vanderplas, A.~Passos,
  D.~Cournapeau, M.~Brucher, M.~Perrot, and E.~Duchesnay.
\newblock Scikit-learn: Machine learning in {P}ython.
\newblock {\em Journal of Machine Learning Research}, 12:2825--2830, 2011.

\bibitem{renard2012}
F.~Renard, C.~Heinrich, S.~Achard, E.~Hirsch, and S.~Kremer.
\newblock Statistical kernel-based modeling of connectomes.
\newblock In {\em Proc. Int Workshop on Pattern Recognition in {NeuroImaging}},
  pages 69--72, 2012.

\bibitem{richiardi2011}
Jonas Richiardi, Sophie Achard, Edward Bullmore, and Dimitri Van~De Ville.
\newblock Classifying connectivity graphs using graph and vertex attributes.
\newblock In {\em Proc. Int Workshop on Pattern Recognition in {NeuroImaging}},
  Seoul, Korea, 2011.

\bibitem{richiardi2013}
Jonas Richiardi, Sophie Achard, Horst Bunke, and Dimitri Van De~Ville.
\newblock Machine learning with brain graphs: predictive modeling approaches
  for functional imaging in systems neuroscience.
\newblock {\em IEEE Signal Processing Magazine}, 30(3):58--70, 2013.

\bibitem{robinson2010}
Emma~C Robinson, Alexander Hammers, Anders Ericsson, A~David Edwards, and
  Daniel Rueckert.
\newblock Identifying population differences in whole-brain structural
  networks: a machine learning approach.
\newblock {\em {NeuroImage}}, 50(3):910--919, April 2010.

\bibitem{rosazza2011resting}
Cristina Rosazza and Ludovico Minati.
\newblock Resting-state brain networks: literature review and clinical
  applications.
\newblock {\em Neurological sciences}, 32(5):773--785, 2011.

\bibitem{sfikas2016}
Giorgos Sfikas and Christophoros Nikou.
\newblock Bayesian multiview manifold learning applied to hippocampus shape and
  clinical score data.
\newblock In {\em Medical Computer Vision and \uppercase{B}ayesian and
  Graphical Models for Biomedical Imaging -- {\sc miccai} 2016}, pages
  160--171. Springer, 2016.

\bibitem{sporns2005}
Olaf Sporns, Giulio Tononi, and Rolf K{\"o}tter.
\newblock The human connectome: a structural description of the human brain.
\newblock {\em PLoS Computational Biology}, 1(4):e42, 2005.

\bibitem{supekar2008}
Kaustubh Supekar, Vinod Menon, Daniel Rubin, Mark Musen, and Michael~D.
  Greicius.
\newblock Network analysis of intrinsic functional brain connectivity in
  alzheimer's disease.
\newblock {\em PLoS Computational Biology}, pages 1--11, June 2008.

\bibitem{tenenbaum2000}
Joshua~B. Tenenbaum, Vin de~Silva, and John~C. Langford.
\newblock A global geometric framework for nonlinear dimensionality reduction.
\newblock {\em Science}, 290(5500):2319 --2323, December 2000.

\bibitem{watts1998}
Duncan~J Watts and Steven~H Strogatz.
\newblock Collective dynamics of ‘small-world’networks.
\newblock {\em nature}, 393(6684):440, 1998.

\bibitem{webb02}
Andrew Webb.
\newblock {\em Statistical pattern recognition}.
\newblock Wiley, second edition, 2002.

\bibitem{wolpert1996}
David~H Wolpert.
\newblock The lack of a priori distinctions between learning algorithms.
\newblock {\em Neural computation}, 8(7):1341--1390, 1996.

\end{thebibliography}
\end{document}